\documentclass[prd,twocolumn,nofootinbib,amssymb, amsmath,mathrsfs,nobibnotes,aps]{revtex4}
\usepackage{color}
\usepackage{graphicx}
\usepackage[colorlinks=true,citecolor=blue,linkcolor=blue]{hyperref}
\usepackage{times}
\usepackage{amsmath}

\newcommand{\lsim}{\lower0.6ex\vbox{\hbox{$ \buildrel{\textstyle <}\over{\sim}\ $}}}
\newcommand{\gsim}{\lower0.6ex\vbox{\hbox{$ \buildrel{\textstyle >}\over{\sim}\ $}}}
\newcommand{\beq}{\begin{equation}}
\newcommand{\eeq}{\end{equation}}

\begin{document}

\title{Bounds on Resonantly-Produced Sterile Neutrinos from Phase Space Densities of Milky Way Dwarf Galaxies
}


\author{Mei-Yu Wang$^{1,2}$\footnote{meiyuw@andrew.cmu.edu}}
\author{John F. Cherry$^{3,4}$}
\author{Shunsaku Horiuchi$^{3,}$}
\author{Louis E. Strigari$^{2,}$}
\affiliation{$^1$Department of Physics, Carnegie Mellon University, Pittsburgh, PA 15213}
\affiliation{$^2$Department of Physics \& Astronomy, Mitchell Institute for Fundamental Physics and Astronomy, Texas A\&M University, College Station, TX 77843-4242}
\affiliation{$^3$Center for Neutrino Physics, Department of Physics, Virginia Tech, Blacksburg, VA 24061, USA}
\affiliation{$^4$Department of Physics, University of South Dakota, Vermillion, SD 57069, USA}


\begin{abstract}
We examine the bounds on resonantly-produced sterile neutrino dark matter from phase-space densities of Milky Way dwarf spheroidal galaxies (dSphs). The bounds result from a derivation of the dark matter coarse-grained phase-space density from the stellar kinematics, which allows us to explore bounds from some of the most compact dSphs without suffering the resolution limitation from N-body simulations that conventional methods have. We find that the strongest constraints come from very compact dSphs, such as Draco II and Segue 1.  We additionally forecast the constraining power of a few dSph candidates that do not yet have associated stellar kinematic data, and show that they can improve the bounds if they are confirmed to be highly dark-matter dominated systems. Our results demonstrate that compact dSphs provide important constraints on sterile neutrino dark matter that are comparable to other methods using as Milky Way satellite counts. In particular, if more compact systems are discovered from current or future surveys such as LSST or HSC, it should be possible to test models that explain the 3.5 keV X-ray line signal with a 7.1 keV sterile neutrino particle decay. 

\end{abstract}
\maketitle

\section{Introduction}
\label{section:introduction}

One of the most promising class of dark matter (DM) candidates are sterile neutrinos. Occurring as a natural extension to the standard model (SM) of particle physics, they predict a range of rich phenomenologies that have been extensively studied \cite{Adhikari:2016bei,Abazajian:2017tcc}. The presence of mixing between SM active neutrinos and sterile neutrinos will partially or completely thermalize the sterile neutrinos and contribute to both the mass density as well as relativistic energy density of the Universe at early times. The keV-scale mass sterile neutrinos with proper flavor mixing with SM neutrinos can be produced via collisional processes in the early Universe to be the DM \cite{Dodelson_Widrow_1994}. However, this Dodelson-Widrow (DW) mechanism produces large particle momenta and has been shown to significantly conflict with small-scale structure formation and X-ray observations (e.g., \cite{Horiuchi:2013noa}). Alternatively, in the presence of a small primordial lepton asymmetry, the sterile neutrinos can be produced resonantly via in-medium neutrino mixing enhancement and yield colder momenta~\cite{Shi_Fuller_1999}.

The resonantly produced sterile neutrinos will have a non-thermal distorted momentum distribution departing from a Fermi-Dirac distribution. Their non-negligible velocity dispersion makes them a type of warm dark matter (WDM) model that can suppress structure formation below the free-streaming scale, which is also a characteristic for sterile neutrinos generated by DW and other mechanisms. Recent works utilize structure formation information from Lyman-$\alpha$ forest and Milky Way/M31 satellite counts to derive tight constraints that either exclude or leave very little parameter space that is still compatible with X-ray observations \cite{Schneider_2016, Cherry_etal17,Perez_etal2017}. However, these different approaches have different systematic uncertainties. For example, the uncertainties in the assumptions of the thermal history of the intergalactic median (IGM) when modeling the Lyman-$\alpha$ forest flux can affect the derived mass limits \cite{Garzilli_etal2015}. On the other hand, the comparison of Milky Way (MW) satellite counts with theoretical predictions will require corrections for survey completeness for uncovered sky area and survey depth limitations \cite{Tollerud_etal08}. Therefore additional constraints using other methods will help to validate these results.

Phase-space density information of sterile neutrinos have been shown to provide robust bounds on DM models~\cite{Tremaine_Gunn_1979}. According to Liouville's theorem for collisionless systems, the distribution function of the particles is time-independent. Therefore, at the central region of galaxies, a theoretical maximum fine-grained phase-space density exists. Phase space mixing, or coarse-graining, can only decrease the phase space density below this fine-grained bound set by the nature of the particles. Previous works which compare the estimated coarse-grained phase-space density of MW dwarf spheroidal galaxies (dSphs) have set the lower limit of the DW sterile neutrino mass to $m_{\nu_s}^{\rm DW} \gsim$2.5 keV \cite{Hogan_etal00, Gorbunov_etal08, Boyarsky_etal09, Horiuchi_etal14b}, equivalent to approximately 0.7 keV for a thermal WDM particle. 

In this paper we examine the bound on resonantly produced sterile neutrino models using MW dSph phase-space density information. We include several newly-discovered satellite galaxies from wide-field optical surveys such as the Dark Energy Survey (DES) and Pan-STARRS that were not considered in previous studies. We also improve the coarse-grained phase space density modeling by performing a detailed stellar kinematic calculation to derive the DM distribution function implied from observations. We show that this improved framework quantifies the systematic uncertainties that is often omitted or not treated well in the past studies. Using our improved constraints we are able to rule out plenty of resonantly produced sterile neutrino parameter space and provide robust tests of these models. 

This paper is organized as follows: in Section~\S~\ref{section:PSD_limits} we describe theoretical arguments of how limits can be derived using MW dSph phase-space densities. In Section~\S~\ref{section:df} the framework of calculating the DM distribution function of the dSphs and the corresponding coarse-grained phase-space density is discussed. In Section ~\S~\ref{section:results} we present the results of the sterile neutrino fine-grained phase space density maximum calculation and also the predicted coarse-grained phase-space density for MW dSphs. We then utilize this information to derive the limits on resonantly produced sterile neutrinos. 
\section{Dark matter phase-space density limits} 
\label{section:PSD_limits}

Liouville's theorem requires that for dissipationless and collisionless particles the phase-space density cannot increase, and its maximum does not change with time. The estimated coarse-grained phase-space density Q should therefore be smaller than the maximal fine-grained phase-space density $q_{\rm max}$. Since $q_{\rm max}$ depends on the primordial DM properties, the inequality Q $<$ $q_{\rm max}$ can be used to derive, e.g., the sterile neutrino DM mass and mixing angle limits.

To begin with, the number density and pressure of the collisionless gas of sterile neutrino dark matter particles are given by,
\beq
n = \frac{g_\nu}{(2\pi)^3}\int f(p)d^3p\, ,
\label{eq:numbdens}
\eeq
\beq
P = \frac{g_\nu}{(2\pi)^3}\int \frac{p^2 f(p)}{3E}d^3p\, ,
\label{eq:pressure}
\eeq
where $g_\nu$ is the number of spin degrees of freedom for a neutrino and $f(p)$ is the distribution function for resonantly produced sterile neutrinos.  These distribution functions are derived using the publicly available code {\tt sterile-dm} \citep{Venumadhav_etal16}. By including previously neglected effects such as the redistribution of lepton asymmetry and the neutrino opacity, as well as a more accurate treatment of the scattering rates through the quark-hadron transition, the authors provide accurate sterile neutrino phase-space densities. Detailed model properties such as the lepton asymmetry values required to produce the correct relic abundance and the average momentum distribution are discussed in Refs.~\cite{Venumadhav_etal16,Cherry_etal17}.

\begin{figure}
\includegraphics[height=7.8 cm]{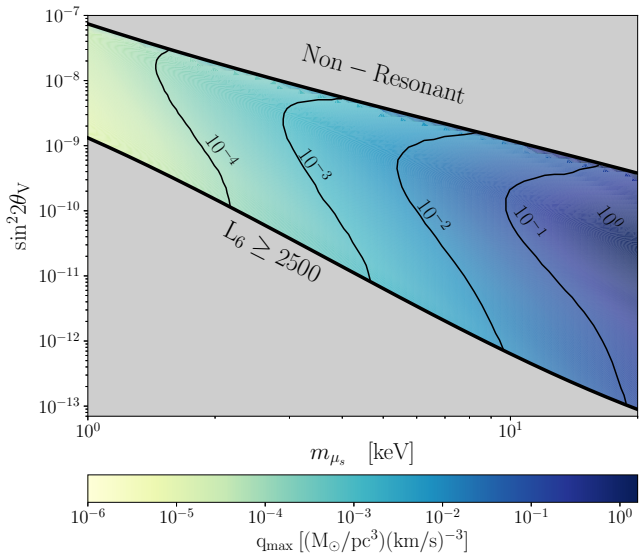}
\vspace{-1.5em}
\caption{Distribution of fine-grained maximal phase-space density ${\rm q_{max}}$ derived from {\tt sterile-dm} \citep{Venumadhav_etal16}. The ${\rm q_{max}}$ values increase from small to high mass, which is similar to the non-resonant case. However the constant contour lines (solid black line) are bent due to the distorted momentum distribution from SM neutrinos mixing.}
\label{fig:qmax}
\end{figure}

We can then write down the primordial phase space density $q_{\rm max} = \rho/\langle v^2\rangle^{3/2}$ of the resonantly produced sterile neutrinos and relate it to our expressions for pressure and number density via the ideal gas law, $\langle v^2 \rangle = 3P/nm_{\nu_s}$.  This leads to the general expression for the fine-grained phase space density for resonantly produced sterile neutrino dark matter,

\beq
q_{\rm max} =  \frac{g_\nu m_{\nu_s}^4}{2\pi^2}\times\frac{\left[\int p^2f(p)dp\right]^{5/2}}{\left[\int p^4f(p)dp\right]^{3/2}}\, .
\label{eq:qmax}
\eeq
The distribution of fine-grained phase space density for resonantly produced sterile neutrinos are shown in Figure~\ref{fig:qmax}.

The coarse-grained phase-space density Q is defined as the mass density in a finite six-dimensional phase-space volume. Dynamically relaxed systems often have a Maxwellian-like velocity distribution, which is also found to be a good description for the central regions of the dark matter halo \cite{Vogelsberger_etal09}. The maximum density in velocity is then $(2\pi\sigma^2)^{-3/2}$ and the corresponding maximum coarse-grained phase space density is:
\beq
Q\equiv {\bar{\rho}\over (2\pi \sigma^2)^{3/2}} 
\label{eq:Q_mb}
\eeq
where $\bar{\rho}$ is the average DM density of the system and $\sigma$ is the one-dimensional DM velocity dispersion. This definition is found to predict the maximum coarse-grained phase space density in N-body simulations remarkably well \cite{Shao_etal2013}, while other definitions, such as the one proposed in \cite{Hogan_etal00}, may overestimate the true phase space density significantly. We therefore adopt the this definition $Q$ in our entire calculation. In the literature, there are several different definitions of Q that have been adopted (e.g. see \cite{Hogan_etal00,Simon_etal07,Boyarsky_etal09}), but the conversion between different forms can be easily made by multiplying by the appropriate constant factors. 

\section{dSph coarse-grained phase-space density}
\label{section:df}
In a dSph, the observable quantities are the projected positions and velocities of stars. The phase-space density distribution of DM can be estimated from these observations. However, often in the literature the direct calculation of the DM velocity distribution is ignored and the results are derived simply assuming that the DM velocity dispersion is similar to the measured stellar velocity dispersion, which is in general not true. One possible approach to tackle this problem is to derive DM phase-space densities from N-body numerical simulations \citep{Shao_etal2013,Horiuchi_etal14b}. However, because we are interested in some of the smallest galaxies in the Universe, these estimates suffer  from simulation resolution limitations as we go to very compact ultra-faint dSphs. 

Here we perform analytical calculations to map the viable DM phase space distribution that satisfies the observed stellar properties in the dSphs. Assuming the system is spherically symmetric and orbits for both stars and DM particles are isotropic, the distribution function follows the Eddington formula for a given potential $\Psi(r) $,
\begin{equation} 
f_{\rm DM/star}(\epsilon) = \frac{1}{\sqrt{8}\pi^2} \int_\epsilon^0 
\frac{d^2 \rho_{\rm DM/star}}{d\Psi^2}\frac{d\Psi}{\sqrt{\Psi -\epsilon}}. 
\label{eq:eddington}
\end{equation}
Here ${\rm \rho_{DM/star}}$ is either the dark matter density profile or the stellar density profile, $\epsilon = v^2/2 + \Psi(r)$ is the energy of a DM particle/star, and $v$ is the modulus of the velocity. The function $\Psi(r) $ is the spherically-symmetric potential that depends on the shape of the dark matter density profile. Here we assume that the system potential is dominated by DM, which is true for the dSphs considered here since they typically have high dynamical mass-to-light ratios of a few tens to a few hundred \cite{McConnachie_12}. 

To describe the shape of the dark matter density distribution, we use two canonical examples: the NFW profile,
\beq
\rho_{\rm NFW} (r) = \rho_s/[(r/r_s) (1+r/r_s)^2], 
\label{eq:density_nfw}
\eeq
and the Burkert profile, 
\beq
\rho_{\rm Burkert} (r) = \rho_b /[(1+ r/r_b) (1+(r/r_b)^2)]. 
\label{eq:density_burkert}
\eeq
The NFW profile is the well-known universal fitting to N-body numerical simulations \citep{Navarro_etal1996}. However, since dSph halo profiles can be modified by galaxy formation physics or microphysics of dark matter (e.g., \citep{Vogelsberger_etal2012}), they may have shallower inner profiles. Therefore we adopt two types of profiles, to bracket the plausible shapes: one is cuspy (NFW) and the other is cored (Burkert). 
  
For the stellar density profile, we use the Plummer profile that is widely adopted in the literature,
\beq
\rho_{\rm star} (r) = \rho_p /(1+(r/r_p)^2)^{5/2}. 
\label{eq:density_plummer}
\eeq

To constrain the allowed range of distribution function $f$, we combine the constraints on the dark matter density profiles derived from N-body simulations and the kinematic properties from observations of the stellar velocity dispersion. Specifically, we construct the velocity dispersion profile of the dark matter and stars as a function of distance from the center of the galaxies $r$ by integrating the velocity distribution function derived from Eq.~(\ref{eq:eddington}),
\beq
\langle \sigma_{\rm DM/star}^2 (r)\rangle = {\int v^4 f_{\rm DM/star}(v,r) dv \over \int v^2 f_{\rm DM/star}(v,r) dv }.
\label{eq:avgsigma} 
\eeq 

\begin{table}
\caption{dSph properties and derived DM velocity dispersion}
{\renewcommand{\arraystretch}{1.5}
\renewcommand{\tabcolsep}{0.15cm}
\centering
\begin{tabular}{l c c c c c c c }
\hline 
\hline
dSph name &  $\langle \sigma_{*} \rangle$ & $\epsilon$ & 2D $r_h$ & 3D $r_h^{\dagger}$ & Ref$.^{\dagger \dagger}$\\
  & [km/s] &  & [pc] &  [pc] &\\
\hline
Coma Berenices&$4.6^{+0.8}_{-0.8}$& 0.38&$77.0$& $78.2$& (1)\\
Pegasus III &$5.4^{+3.0}_{-2.5}$& 0.38&$53.0$& $54.3$&(2) (3)\\
Horologium I&$4.9^{+2.8}_{-0.9}$& --- &$60.0$& $78.0$& (4) (5) (6)\\
Reticulum II&$3.3^{+0.7}_{-0.7}$& 0.6&$55.0$& $45.2$& (4) (6) (7)\\
Segue 1 &$3.9^{+0.8}_{-0.8}$& 0.48&$29.0$& 27.2&  (1)\\
Draco II&$2.9^{+2.1}_{-2.1}$&0.24 &$19.0$& $21.5$& (8) (9)\\
\hline
Cetus II& --- & 0.4&$17.0 $& $17.12$& (10)\\
Tucana V&--- & 0.7&$17.0 $& $12.10$&  (10)\\
\hline
\hline
\end{tabular}\\
\vspace{-0.7em}
\begin{flushleft}
$\dagger$ Azimuthal 3D $r_h$ = 1.3 $\times$ 2D $r_h \times \sqrt{1-\epsilon}$\\
$\dagger \dagger$ References abbreviated as : (1) \citet{McConnachie_12}, (2) \citet{Kim_etal16}, (3) \citet{Kim_etal15b}, (4) \citet{Bechtol_etal15} (5) \citet{Koposov_etal15b}, (6) \citet{Koposov_etal15}, (7) \citet{Simon_etal15}, (8) \citet{Martin_etal16}, (9) \citet{Laevens_etal15}, (10) \citet{Drlica_Wagner_etal15} \\
\end{flushleft}
 }
\label{tb:dSphs}
\end{table}

\begin{figure*}
\includegraphics[height=8.5 cm]{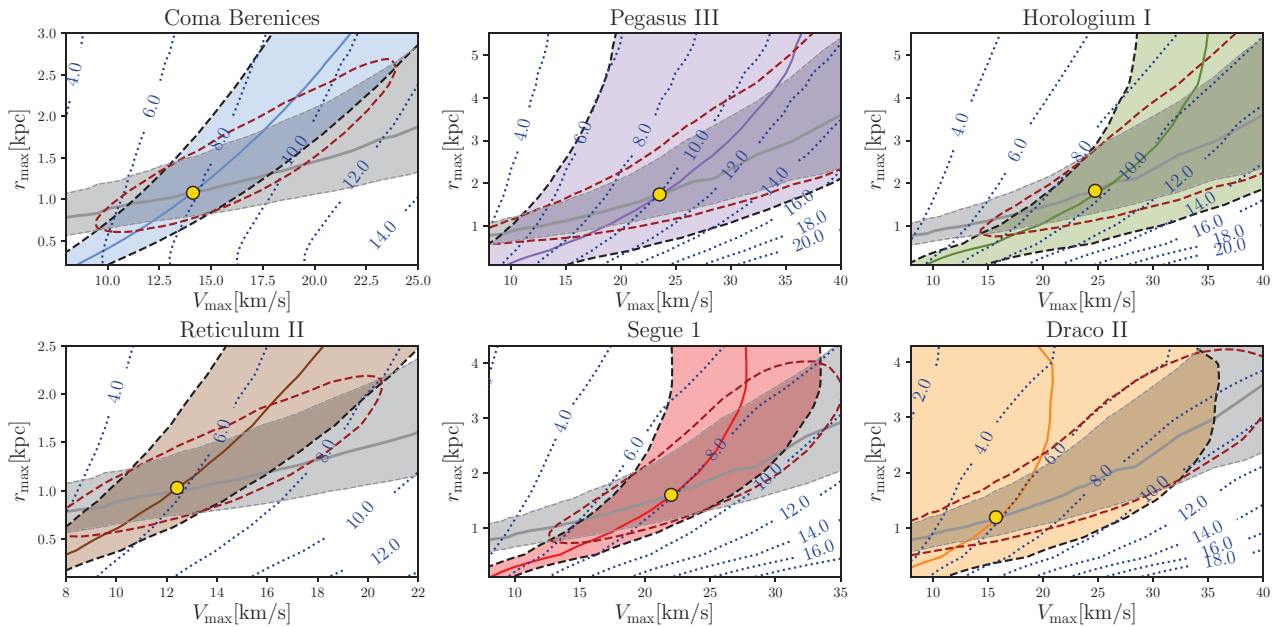}
\vspace{-1em}
\caption{Contours of dark matter velocity dispersion (blue dotted lines) as a function NFW profile parameter $V_{\rm max}$ and $r_{\rm max}$ for six Milky Way dSphs. The gray solid lines and gray shaded regions show profile parameter regions that agree with the subhalo $V_{\rm max}$ - $r_{\rm max}$ relation from COCO-COLD simulations (median values and enclosed 16th and 84th percentile regions). The colored regions encompass the regions where the profiles predict dynamical mass that agree with the measured stellar velocity dispersion within 1$\sigma$ uncertainty values. The solid color lines represent the models that agree with the median stellar velocity dispersion values. The yellow circle points mark the the best-fit values for from the joined likelihood functions of stellar velocity dispersion and N-body simulations, and the dark red dash lines show the enclosed 1 $\sigma$ uncertainty regions.}
\label{fig:dSph_NFW}
\end{figure*}

The average velocity dispersions, for either stars or DM particles, for each DM potential assumption for a given system, are then derived by weighted averaging over the DM or star density profile, $\rho_{\rm DM/star}$, within the 3D azimuthal half-light radius. We then select the viable range of DM profile parameters using a combined 2D Gaussian likelihood function constructed from the stellar velocity dispersion and DM density profile properties from N-body simulations. Instead of using parameters $(\rho_s,r_s)$ and $(\rho_b,r_b)$ shown in Eq.~(\ref{eq:density_nfw}) and Eq.~(\ref{eq:density_burkert}), we display the profile parameters in Section ~\S~\ref{section:results} as ${\rm V_{max}}$ and ${\rm r_{max}}$. The relation between these parameters are,
\beq
{\rm r_{max}}=2.16 r_s \; {\rm or} \; 3.28 r_b,  
\label{eq:rmax_vmax}
\eeq
and ${\rm V_{max}}$ is the maximum of the circular velocity ${\rm \sqrt{GM(<r)/r}}$, where ${\rm M(<r)}$ is the mass enclosed within radius r. For each part, either the stellar velocity dispersion or N-body simulation results, we assume they are 2D Gaussians with 1 $\sigma$ error width from observations (see Table~\ref{tb:dSphs} for stellar velocity dispersion uncertainties) or the corresponding 16th and 84th percentile regions (for N-body simulation results). We then multiply the two likelihood functions assuming no correlation to derive the joined likelihood function. The best-fit profile parameter values are selected with the highest likelihood values with 1 $\sigma$ uncertainty ranges. The coarse-grained phase space density for each dSph is then derived by combining the averaged DM density and DM velocity dispersion within the 3D azimuthal half-light radius (see Eq.~\ref{eq:Q_mb}). The average density is calculated using the analytical density form from Eqs.~(\ref{eq:density_nfw}) or (\ref{eq:density_burkert}). 

Here we consider several compact dSphs, and many of them are newly discovered ultra-faints or ultra-faint candidates: Coma Berenices, Pegasus III, Horologium I, Reticulum II, Segue 1, Draco II, Cetus II, and Tucana V. Their properties such as stellar velocity dispersion, ellipticity, and 2D and 3D azimuthal half-light radius are listed in Table~\ref{tb:dSphs}. These systems have low velocity dispersion and high DM density within the galaxies. Therefore they yield large phase-space density values and are ideal targets for studying sterile neutrino limits. Here Cetus II and Tucana V are ultra-faint candidates that have no spectrascopic follow-up yet.  Among these dSphs, other than Coma Berenices, Segue 1, and Pegasus III~\citep{Kim_etal15b} that are found using Sloan Digital Sky Survey (SDSS) data, the rest are found recently from the Dark Energy Survey (DES) \citep{Bechtol_etal15, Drlica_Wagner_etal15,Koposov_etal15b} (Horologium I, Reticulum II, Cetus II, and Tucana V) and Pan-STARRS \citep{Laevens_etal15} (Draco II) data. However, we note that \cite{Conn_etal2017} had acquired deep Gemini/GMOS-S to study three ultra-faint dwarf galaxy candidates including Tucana V. They argue that Tucana V has low-level excess of stars in their data without a well-defined centre, and it is likely either a chance grouping of stars related to the Small Magellanic Cloud (SMC) halo or a star cluster in an advanced stage of dissolution.

\section{Results}
\label{section:results}

\subsection{Coarse-grained phase space density of dSphs}
\label{subsection:coarse-grained}
In this section we present results for the allowed coarse-grained phase-space density of dSphs by combining stellar kinematic information and N-body numerical simulation predictions. 

In Figure~\ref{fig:dSph_NFW} we explore the viable regions in terms of DM density profile parameters ${\rm V_{max}}$ and ${\rm r_{max}}$ for NFW profiles and the corresponding predictions for the DM velocity dispersions (color contours). Here we restrict the ${\rm V_{max}}$ to be in the range 8--40 km/s, which is supported by recent N-body simulations of MW dSphs (e.g. \cite{Fattahi_etal16,Fitts_etal16}). The lower limit of ${\rm V_{max}} >$ 8 km/s comes from the argument that the UV background from re-ionization suppresses star formation below certain halo mass threshold~\citep{Okamoto_etal08}.
 
The colored regions in Figure~\ref{fig:dSph_NFW} map the regions where the derived average stellar velocity dispersion values using Eq.~(\ref{eq:avgsigma}) agree with the measured values within the 1$\sigma$ uncertainties. The solid color lines show results that match the median values, and the shaded region encompass the 1$\sigma$ uncertainty regions listed in column 2 of Table~\ref{tb:dSphs}. The solid gray curves indicate the median ${\rm V_{max}}$ - ${\rm r_{max}}$ relation for subhalos in the DM-only COCO-COLD simulations~\cite{Bose_etal2017} (in their Figure 9, middle lower panel with $M_{200}^{\rm host} = 1-4\times 10^{12} h^{-1} {\rm M}_{\odot}$) and the gray shaded region show their 16th and 84th percentile range. Assuming both the stellar velocity dispersion and the ${\rm V_{max}}$ - ${\rm r_{max}}$ relation from simulations can be well approximated by 2D Gaussian functions in the profile parameter space, the yellow points in each panel mark the the best-fit values from the joint likelihood function and the dark red dash lines show the enclosed 1$\sigma$ uncertainty regions. The corresponding range of DM velocity dispersion and Q derived from the NFW (Burkert) profiles are listed in Table~\ref{tb:Q_value}.

\begin{figure*}
\includegraphics[height=8.5 cm]{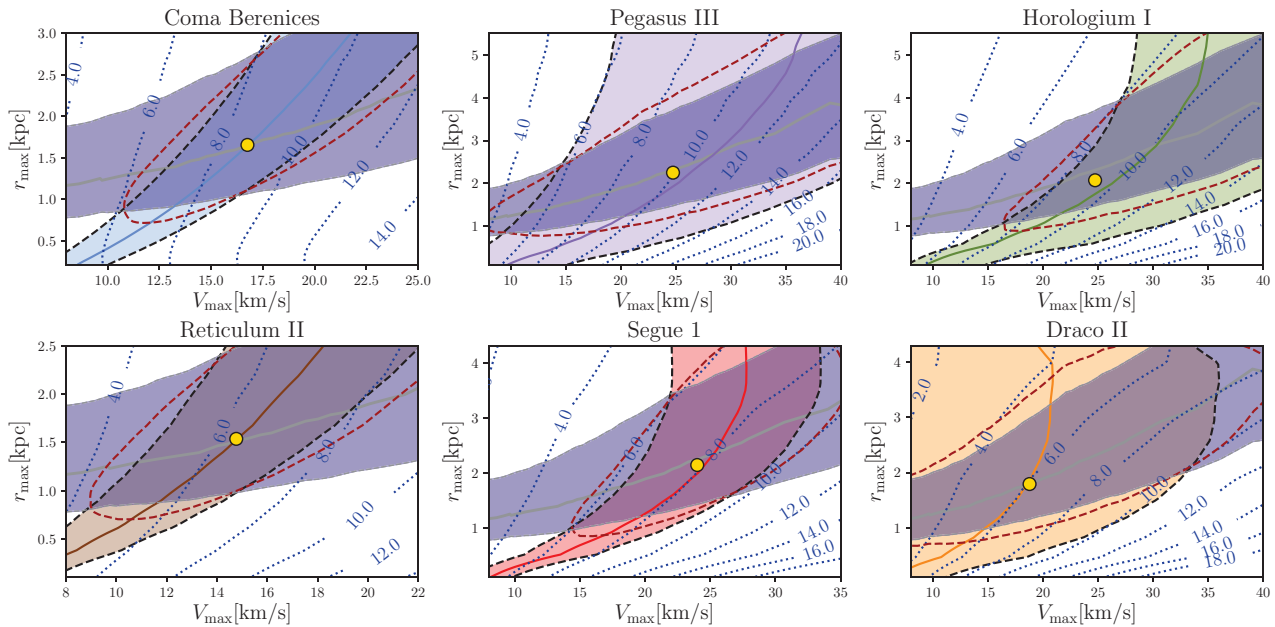}
\vspace{-1em}
\caption{The same as Figure~\ref{fig:dSph_NFW} except that the dark blue solid lines and shaded regions show profile parameter regions that agree with the subhalo $V_{\rm max}$ - $r_{\rm max}$ relation from COCO-WARM simulations (median values and enclosed 16th and 84th percentile regions). }
\label{fig:dSph_NFW_warm}
\end{figure*}

\begin{figure}
\includegraphics[height=11.0 cm]{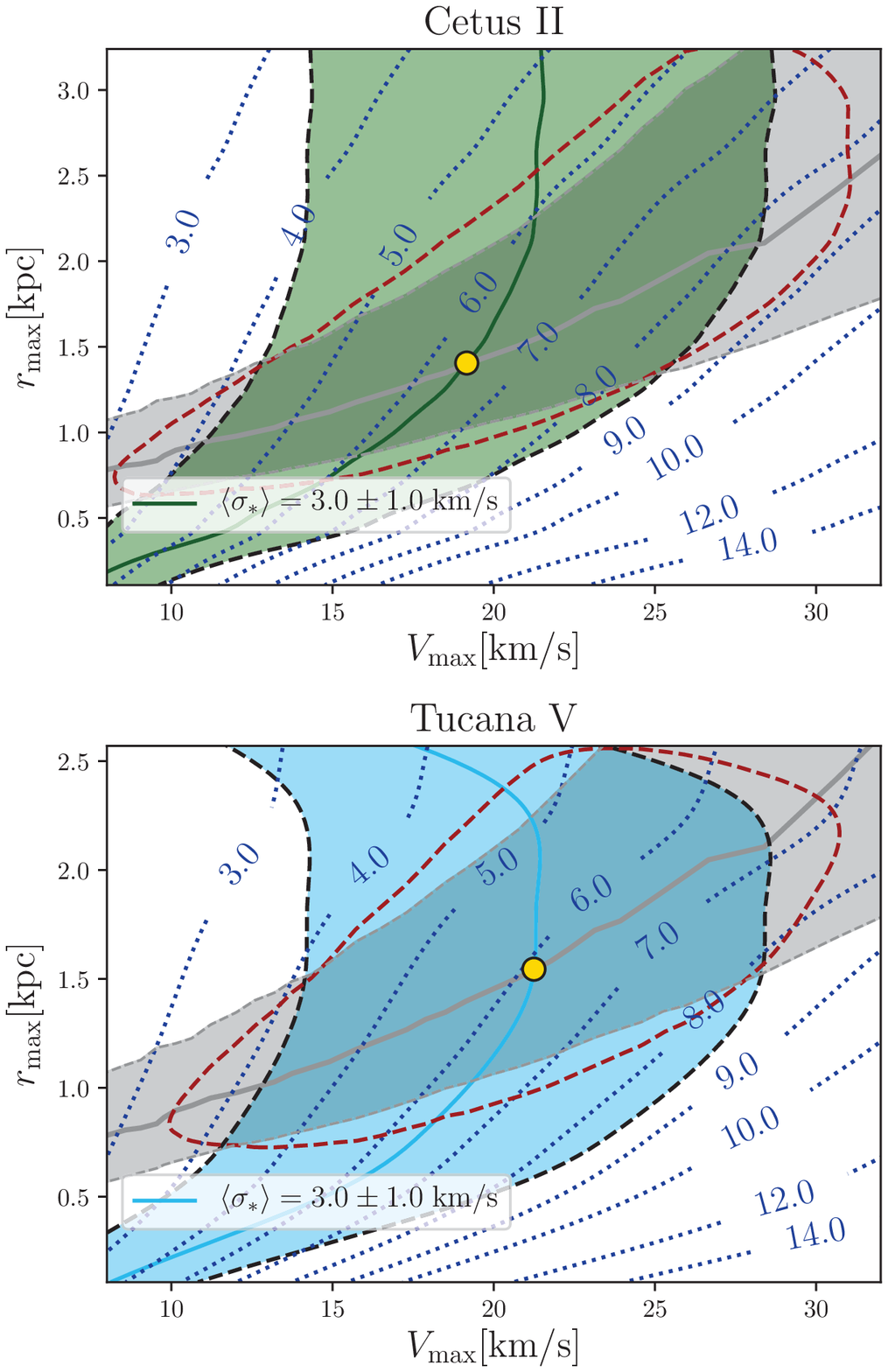}
\caption{Contours of projected DM velocity dispersion values (blue dotted lines) as a function of NFW profile parameter $V_{\rm max}$ and $r_{\rm max}$ for two dSph candidates, Cetus II and Tucana V. The gray bands show the subhalo $V_{\rm max}$ - $r_{\rm max}$ relation from the COCO-COLD simulations \citep{Bose_etal2017}. The solid color lines show the median stellar velocity dispersion values of 3.0 km/s, and the color contours span the regions assuming velocity uncertainties of 1.0 km/s. Again the yellow points mark the the best-fit values for from the joined likelihood functions from stellar velocity dispersion and N-body simulations, and the dark red dash lines show the enclosed 1 $\sigma$ uncertainty regions.}
\label{fig:prediction}
\end{figure}

We note that our stellar kinematic calculation results using Burkert profiles have little to no overlap with the N-body simulation predictions. To generate the dynamical mass that matches the observed stellar kinematics, the Burkert profiles are required to have much smaller characteristic radius $r_b$, therefore ${\rm r_{max}}$, than NFW profiles, which is cuspy at the central region. However, as we show in Figure~\ref{fig:dSph_NFW_warm}, the ${\rm V_{max}}$ - ${\rm r_{max}}$ relation for shallower profiles, such from a warm dark matter (WDM) simulation (blue shaded contour and lines), occupies regions with even higher ${\rm r_{max}}$ than the corresponding CDM model (gray shaded contours and lines in Figure~\ref{fig:dSph_NFW}). Here the WDM ${\rm V_{max}}$ - ${\rm r_{max}}$ relation is from the COCO-WARM simulation ~\cite{Bose_etal2017}, which is also from the middle lower panel of their Figure 9. These authors have adopted a model with similar cutoff on the matter power spectrum as a 7 keV sterile neutrino with leptogenesis parameter $L_6 \sim$ 8.66, which is a good benchmark model for our work. Therefore in our studies the dynamical mass predictions using Burkert profiles are often in tension with the N-body simulation results, and we conclude that Burkert profile predictions are in general not good fits to most of the compact dSphs we consider here. Therefore, although we list both the derived DM velocity dispersion and coarse-grained Q values for NFW and Burkert assumptions, we only show NFW results for our sterile neutrino model constraints. Also for the Burkert profile results, we only list the possible range of values instead of the best-fit values with 1$\sigma$ uncertainties, as we derived using NFW profile with 2D Gaussian likelihood functions.

\begin{figure}[!t]
\includegraphics[height=6.2 cm]{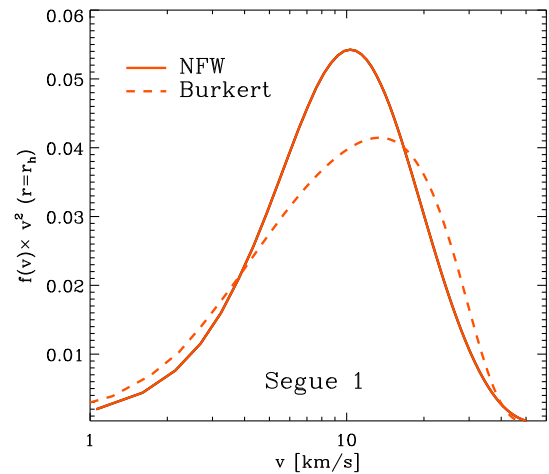}
\vspace{-1em}
\caption{Example of DM velocity distribution function at the half-light radius $r_h$ for Segue 1 for NFW profile (solid line) and Burkert profile (dash line). The profile parameters $V_{\rm max}$ for both NFW and Burkert are fixed at 18 km/s. The $r_{\rm max}$ for NFW is 0.9 kpc and $r_{\rm max}$ = 0.26 kpc for Burkert profiles. Those choices correspond to parameters having similar dynamical mass predictions from the observed average stellar velocity dispersion value. The one with Burkert profile has wider width, which corresponds to higher predicted DM velocity dispersion, than the one with NFW profile.}
\label{fig:Segue_fv}
\end{figure}

In Table~\ref{tb:Q_value} we list the DM velocity dispersion and Q value predictions for several MW dSphs. The derived DM velocity dispersions, both for NFW and Burkert profiles, are larger than the measured stellar velocity dispersion values, which are shown in Table~\ref{tb:dSphs}. The DM velocity dispersions for NFW profiles are smaller for galaxies with small stellar velocity dispersion, although a similar trend doesn't appear on the Burkert profile predictions. Nevertheless, the Q value predictions show a consistent trend of increasing for smaller galaxies with lower stellar velocity dispersion. The DM velocity dispersions for Burkert profiles $\langle \sigma_{\rm DM}^{\rm Burkert} \rangle$ are in general larger than the NFW predictions $\langle \sigma_{\rm DM}^{\rm NFW} \rangle$, which is demonstrated in Figure~\ref{fig:Segue_fv} in terms of the width of the velocity distribution function. On the other hand, the predicted Q values for Burkert profiles, $Q_{\rm \sigma_{\rm DM}}^{\rm Burkert}$, are often smaller than the NFW ones. Since the dynamical mass, which is directly related to the estimated average density $\bar{\rho}$ of the systems, is well constrained by stellar kinematics, the Q value is therefore strongly dependent on the DM velocity dispersions (see Eq.~\ref{eq:Q_mb}), which is proportional to $\sigma^{-3}$, resulting in different Q values for the NFW and Burkert profiles. 
Our results also indicate that wrongly assuming that the DM and stellar velocity dispersion values are similar  result in overestimations of Q and therefore overestimations of the sterile neutrino mass constraints. A simple example can be derived if we assume that the distribution function is Fermi-Dirac-like, which is true for many non-resonant production mechanisms. In this scenario the $q_{\rm max}$ is proportional to ${\rm m_{\nu_s}^4}$, and therefore the mass bound is proportional to $\sigma^{-3/4}$. If the Segue 1 mass bound is derived using $\sigma_*$ = 3.9 km/s instead of $\sigma$ =8.2 km/s, then the 52$\%$ velocity dispersion underestimate will translate into a $\sim$ 75$\%$ mass limit overestimation. 

In Table~\ref{tb:Q_value} we also show predictions of two dSph candidates that do not yet have associated stellar kinematic measurements: Cetus II and Tucana V. We list predictions for Q and $\sigma$ values if they have $\sigma_*$ = 3.0 $\pm$ 1.0 km/s. The predicted dark matter velocity dispersion as a function of density profile parameter are shown in Figure~\ref{fig:prediction}. This choice of $\sigma_*$ range is motivated by previous studies~\cite{Walker_etal09}, which show that the Local Group dSph data exhibit a correlation between stellar velocity dispersion and half-light radius. This results in a scaling relation for dynamical mass within $r_h$, $M(< r_h) \propto r_h^{1.4}$. If we extrapolate this relationship, the predicted stellar velocity dispersion values for Cetus II and Tucana V are about 3-4 km/s, which is comparable to other ultra-faint dSphs with similar $r_h$ such as Segue 1 and Draco II. The expected velocity uncertainties are estimated to be comparable or worse than Reticulum II or Segue 1 because it is often difficult to get substantial stellar kinematic sample from very faint objects. However, we note that for Draco II the velocity uncertainty is particularly large due to its high velocity measurement systematics ($\sim$ 2.25 km/s for DEIMOS) and small sample (9 member stars) \cite{Martin_etal16} that make resolving the intrinsic stellar dispersion difficult. If the true measured $\sigma_*$ is lower than 3 km/s, the corresponding Q values will increase partially due to the decreasing $\sigma$.

\begin{table*}
\caption{DM velocity dispersion and coarse-grained phase space density values (Q) for dSphs with NFW or Burkert profile. }
{\renewcommand{\arraystretch}{1.15}
\renewcommand{\tabcolsep}{0.15cm}
\begin{tabular}{l c c c c c}
\hline 
\hline
dSph name &  $\langle \sigma_{\rm DM}^{\rm NFW} \rangle$&$\langle \sigma_{\rm DM}^{\rm Burkert} \rangle^{\dagger}$&$Q_{\rm \sigma_{\rm DM}}^{\rm NFW} $& ${Q_{\rm \sigma_{\rm DM}}^{\rm Burkert}}^{\dagger} $ \\
  &  [km/s] &[km/s] &$[10^{-5} ({\rm M_{\odot}/pc^3})(\rm{km/s})^{-3}]$& $[10^{-5} ({\rm M_{\odot}/pc^3})(\rm{km/s})^{-3}]$ \\
\hline
Coma Berenices&$7.8^{+ 3.5}_{-2.1}$ & 5.0 -- 22.3&$6.9^{+ 5.6}_{-3.8}$ & 0.2 -- 30.6 \\
Pegasus III & $10.2^{+ 5.7}_{-6.1}$ &5.0 -- 22.1& $5.9^{+ 14.6}_{-2.7}$ & 0.1-- 52.0 \\
Horologium I&$9.7^{+ 4.6}_{-2.9}$ &6.8 -- 23.0& $15.8^{+ 13.3}_{-6.6}$ & 0.8 -- 62.5 \\
Reticulum II&$5.9^{+ 2.4}_{-1.9}$ & 4.9 -- 22.0 &$25.8^{+ 18.3}_{-11.7}$ & 0.1 -- 65.2 \\
Segue 1& $8.2^{+ 2.2}_{-2.6}$ &4.1 -- 22.4& $34.0^{+ 30.9}_{-12.4}$ & 0.7 -- 96.8  \\
Draco II& $6.2^{+ 4.4}_{-2.9}$ & 4.0 -- 21.8& $74.8^{+ 90.9}_{-40.3}$ & 0.1 -- 123.3 \\
\hline
Cetus II *&$6.4^{+ 2.4}_{-3.1}$ & 5.6 -- 21.8 &$100.4^{+ 147.5}_{-36.6}$ &0.1 -- 151.6  & \\
Tucana V *&$6.1^{+2.0}_{-2.6}$ & 5.5 -- 21.8 &$185.4^{+ 194.5}_{-45.5}$ & 0.1 -- 136.9 &\\
\hline
\end{tabular}\\
\smallskip
$*$ We had assumed the measured stellar velocity dispersion is 3.0 $\pm$ 1.0 km/s for both Cetus II and Tucana V.\\
$^{\dagger}$ We only show the possible range of $\langle \sigma_{\rm DM}^{\rm Burkert} \rangle$ and $Q_{\rm \sigma_{\rm DM}}^{\rm Burkert}$ instead of 1 $\sigma$ uncertainties.
 }
\label{tb:Q_value}
\end{table*}

\begin{figure*}
\includegraphics[height=7.5 cm]{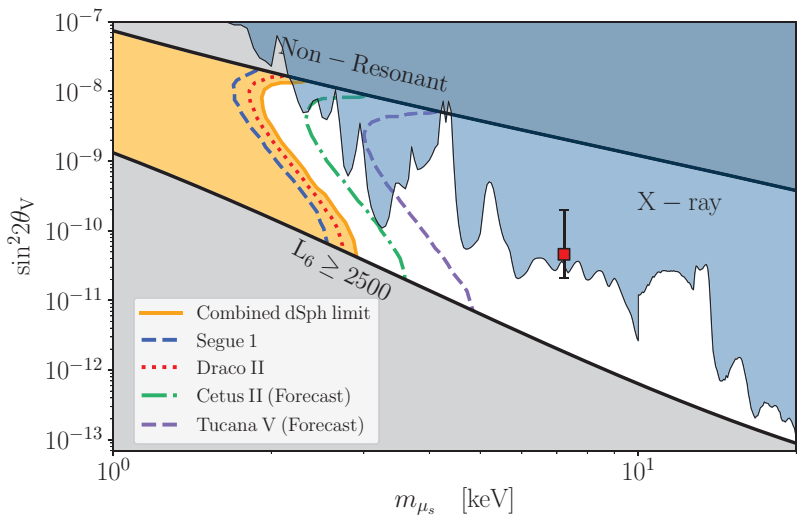}
\vspace{-2.0em}
\caption{Milky Way dSph phase-space density constraints on resonantly produced sterile neutrinos assuming NFW DM profiles. The yellow contour region indicates the currently possible 95$\%$ C.I. exclusion from combining dSphs with stellar kinematic measurement. The red dotted line shows limit from Draco II which provide the best single object constraints. Limit from Segue 1 is shown as the blue dash line. Other two lines indicate forecasts of dSph candidate constraints: for Cetus II (green dash dot line), Tucana V (purple dash line), all assuming stellar velocity dispersion for both Cetus II and Tucana V to be 3.0$\pm$1.0 km/s. The regions in steel blue show the combined 95$\%$ C.I. X-ray exclusion based on deep Chandra observation of M31 and NuSTAR observations of the galactic center \citep{Horiuchi_etal14b,Perez_etal16}. The red rectangular point marks the claimed 7.1 keV sterile neutrino decay line candidate \citep{Boyarsky_etal2014,Bulbul_etal14}. }
\label{fig:constraints}
\end{figure*}

\subsection{Sterile Neutrino DM constraints}
\label{subsection:SN}

We now place constraints on resonantly-produced sterile neutrinos using the Q values derived above. The limits are derived from the fact that the coarse-grained phase space density Q should be smaller than the maximal fine-grained phase-space density $q_{\rm max}$. From the errors on Q values in Table~\ref{tb:Q_value}, we derive the one-sided 95$\%$ confidence interval (C.I.) mass limits (as a function of mixing angle) for both individual galaxies and combined galaxy limits. 

A selection of the resulting limits are shown in Figure~\ref{fig:constraints}. In this figure, the upper thick black line shows the non-resonant DW production parameter region, above which the sterile neutrino budget would overclose the universe (upper light gray area). The lower thick black line corresponds to the resonant production using the maximum lepton asymmetry allowed by Big-bang nucleosynthesis, below which not enough sterile neutrinos would be generated to match the DM budget (lower light gray area). 

We show the constraints for combining dSphs (in yellow) that currently have stellar velocity measurements and from two dSph (Draco II in red dotted line and Segue 1 in blue dash line) that possess the largest Q values. We note that here we only show results from our predictions using NFW profiles (4th column in Table~\ref{tb:Q_value}). The combined limits exclude sterile neutrino mass $m_{\nu_s} \sim$ 1.9--2.8 keV for mixing angle ${\rm sin^2 2\theta_V \sim }$ 1.4$\times 10^{-8}$--4.4 $\times 10^{-11}$. The constraining power comes mainly from two dSphs: Draco II and Segue 1. The limits from Draco II, which provide the best single object limits, exclude sterile neutrino mass $m_{\nu_s} \sim$ 1.8--2.7 keV for mixing angle ${\rm sin^2 2\theta_V \sim }$ 1.8$\times 10^{-8}$--5.7 $\times 10^{-11}$. Segue 1, which is shown as the blue dash line in Figure~\ref{fig:constraints}, provides the second-best limits among our sample of galaxies. It can exclude sterile neutrino mass $m_{\nu_s} \sim$ 1.7--2.5 keV for mixing angle ${\rm sin^2 2\theta_V \sim }$ 2.0$\times 10^{-8}$--7.1 $\times 10^{-11}$. 

We also show forecasts for the limits that can be obtained if two compact dSph candidates (Cetus II in green dash dot line; Tucana V in purple dash line) are confirmed to possess large DM distributions with $\sigma_*$ = 3.0 $\pm$ 1.0 km/s. Tucana V can exclude the parameter space of $m_{\nu_s} \sim$ 3.0--4.9 keV for mixing angle ${\rm sin^2 2\theta_V \sim }$ 5.2$\times 10^{-9}$--7.4 $\times 10^{-12}$. For Cetus II it can exclude a parameter space of $m_{\nu_s} \sim$ 2.3--3.6 keV for mixing angle ${\rm sin^2 2\theta_V \sim }$ 9.0$\times 10^{-9}$--2.1 $\times 10^{-11}$. Several claimed detection signals from X-ray observation of possible sterile neutrino decay (e.g. \citep{Boyarsky_etal2014,Bulbul_etal14}), which suggest $m_{\nu_s} \sim$ 7.1 keV with ${\rm sin^2 2\theta_V \sim }$ 4.6--6.8 $\times 10^{-12}$, is also shown in red square point in Figure~\ref{fig:constraints} for comparison.


\section{Conclusion}
\label{section:conclusion}

In this work we derive the limits on resonantly produced sterile neutrino models using phase space density information from compact MW dSphs. Utilizing the fact that for a systems consisting of dissipationless and collisionless particles the phase-space density cannot increase, limits on sterile neutrino properties can be derived from the inequality that the coarse-grained Q values should be smaller than the maximal fine-grained phase space density $ q_{\rm max}$. We implement stellar kinematic modeling to derive the coarse-grained DM phase space density for each system and explore the effects from different DM potentials generated from two profile considerations: NFW and Burkert. In general the dynamical mass predicted by the Burkert profiles doesn't agree with those from the subhalos in N-body numerical simulations, while the NFW profiles show reasonable fits to the data. 

Our calculations show that the derived DM velocity dispersions are usually bigger than the observed stellar dispersion values, by some factor of 1.7--2 typically in the case of NFW density profiles and could be larger in the case of shallower profiles like Burkert. In the NFW cases they gradually decrease for more compact and low stellar velocity dispersion galaxies. Combining with the high average DM density, the estimated coarse-grained phase space density values Q are higher for smaller and lower velocity dispersion systems. We find that among those dSphs that currently have stellar kinematic measurements, Segue 1 and Draco II generate the highest Q values, and therefore provide the best constraints on sterile neutrino parameter space. 

The best limits from Draco II are excluding sterile neutrino mass $m_{\nu_s} \sim$ 1.8--2.7 keV for mixing angle ${\rm sin^2 2\theta_V \sim }$1.8$\times 10^{-8}$--5.7$\times 10^{-11}$, and Segue 1 has slightly worse but similar excluding limit. The combined limits from all dSph can exclude sterile neutrino mass $m_{\nu_s} \sim$ 1.9$-$ 2.8 keV for mixing angle ${\rm sin^2 2\theta_V \sim }$1.4$\times 10^{-8}$--4.4$\times 10^{-11}$. We further demonstrate that more compact dSph candidates, such as Cetus II and Tucana V, can potentially provide better limits if they are confirmed to be DM dominated systems. We also address one source of systematic uncertainty that arise from the underlying assumptions of the DM distribution, which result in uncertainties of the predicted DM velocity dispersion values. This effect is often ignore or not well-treated in the literature or simply assumed that the DM velocity distribution is similar to stellar velocity distribution. 

Our work provides limits that are comparable to other current limits using different means (e.g. MW/M31 satellite counts, Lyman-alpha forest, see \citep{Schneider_2016, Cherry_etal17}). Here we also present a framework to establish the DM phase-space density of compact dSphs that is free from the N-body simulation resolution limitations that previous analysis may suffer from (e.g. \cite{Horiuchi_etal14b}). Although our results are slightly less sensitive comparing to other means and have not yet reached the X-ray signal from testing 7.1 keV sterile neutrino decay scenarios, future and ongoing surveys such as LSST and HSC survey may discover many more compact systems that can improve the limits.  


%
%
\section*{Acknowledgments}
We would like to thank Matthew Walker for helpful discussions. MYW acknowledges support of the McWilliams Postdoctoral Fellowship. SH is supported by the U.S.~Department of Energy under award number de-sc0018327, and LES is supported by the U.S.~Department of Energy award de-sc0010813.


\bibliography{Res-sn_psd}


\end{document}